\documentclass[10pt]{sig-alternate-05-2015}

\usepackage{epstopdf}
\DeclareGraphicsExtensions{.eps}
\usepackage{amsmath}
\usepackage{color}
\usepackage{float}
\usepackage{slashbox}
\usepackage{multirow, array}
\usepackage{enumitem}
\usepackage{wrapfig}
\usepackage{subcaption}
\graphicspath{{figures/}}
\newcommand{\todo}[1]{\textcolor{red}{#1}}
\usepackage{flushend}

\begin{document}
  \CopyrightYear{2016}
  \setcopyright{acmcopyright}
  \conferenceinfo{e-Energy'16,}{June 21--24, 2016, Waterloo, ON, Canada}
  \isbn{978-1-4503-4393-0/16/06}
  \acmPrice{\$15.00}
  \doi{http://dx.doi.org/10.1145/2934328.2934346}






%

\title{Leveraging energy storage to optimize data center electricity cost in emerging power markets}

\numberofauthors{4} 
%
\author{
%
%
\alignauthor
Yuanyuan Shi\\
       \affaddr{Electrical Engineering}\\
       \affaddr{University of Washington}\\
       \affaddr{Seattle, WA}\\
       \email{yyshi@uw.edu}
\alignauthor
Bolun Xu\\
\affaddr{Electrical Engineering}\\
\affaddr{University of Washington}\\
\affaddr{Seattle, WA}\\
\email{xubolun@uw.edu}
\alignauthor Baosen Zhang\\
\affaddr{Electrical Engineering}\\
\affaddr{University of Washington}\\
\affaddr{Seattle, WA}\\
\email{zhangbao@uw.edu}
\and
\alignauthor Di Wang\\
       \affaddr{Microsoft Research}\\
       \affaddr{Redmond, WA}\\
       \email{wangdi@microsoft.com}
}

\maketitle
\begin{abstract}
Energy storage in data centers has mainly been used as devices to backup generators during power outages. Recently, there has been a growing interest in using energy storage devices to actively shape power consumption in data centers to reduce their skyrocketing electricity bills. In this paper, we consider using energy storage in data centers for two applications in a joint fashion: reducing peak demand charges and enabling data centers to participate in regulation markets. We develop an optimization framework that captures the cost of electricity, degradation of energy storage devices, as well as the benefit from regulation markets. Under this framework, using real data Microsoft data center traces and PJM regulation signals, we show the electricity bill of a data center can be reduced by up to 20\%. Furthermore, we demonstrate that the saving from joint optimization can be even larger than the \emph{sum} of individually optimizing each component. We quantify the particular aspects of data center load profiles that lead to this \emph{superlinear} gain. Compared to prior works that consider using energy storage devices for each single application alone, our results suggest that energy storage in data centers can have much larger impacts than previously thought possible. 
\end{abstract}

%
%


%
%

%
%



\section{Introduction}
Management of data center power consumption is becoming increasingly important as data centers grow both in quantity and size. The electricity bill of a data center represents a significant portion of its annual operational cost and can easily reach millions of dollars~\cite{Barroso09}. Consequently, reducing the electricity cost of data centers has received much attention from researchers. 

The electricity bill of a data center is often divided into two parts: energy cost and peak demand charge. The former accounts for the total amount of \emph{energy} consumption of data centers over a certain amount of time (usually one month). The latter is based on the maximum \emph{power} consumption of a data center during that time period. Since a large fraction of capital cost of electrical distribution systems results from accommodating the maximum power draw, the peak demand charge could be as large as the energy cost. Therefore smoothing or flattening peak demands of data centers represents an important method of reducing their electrical bills. 

In addition to cost reduction, data centers can earn revenue by actively participating in the electricity market. Since as the amount intermittent and uncertain renewable generation grows in the power system, more flexible resources are required to maintain balance between supply and demand. In many regions of North America, e.g. California or the PJM control area, this balance is achieved through regulation markets. If the flexibility of data centers can be leveraged successfully, they could play a vital role in integrating renewable resources as well as earning revenue for their owners~\cite{WiermanLiuLiuEtAl2014}. 

 

In this paper we investigate how data centers can shave their peak demands and participate in regulation markets via their batteries. In almost all data centers, batteries normally serve as short term backup in between the failure of outside grid and starting of backup fossil fuel (e.g. diesel) generators. By design, these batteries are sized to be able to power the maximum data center load but are usually never/rarely used in practice~\cite{WangEtAl2012}. Therefore they are natural candidates to provide peak shaving or participate in regulation markets. 

Using batteries for either of the applications is not new, as researchers have considered using them for peak shaving in data centers and much attention have been paid to use energy storage for various applications in the grid. However, both applications have not been \emph{jointly} considered before. In this paper, we demonstrate that significant profit can be gained by performing both application together. Surprisingly, using real data from a Microsoft data center, we show that the joint profit of optimizing for both applications is larger than the \emph{sum} of profits when each application is considered separately.

We demonstrate the reason behind the \emph{superlinear} gain through detailed analysis of both synthetic and actual data center load traces. Intuitively, this gain occurs because of the particular type of batteries used in data centers, as well as the nature of data center load profiles and  regulation signals. The batteries in data centers are used to pick up the entire load of  data centers for a short time before backup generation kicks in, which means that they have high power density but low energy capacity. Because of the fairly broad peaks in load profile, the low energy capacity in these batteries limit their ability in providing peak shaving. On the other hand, regulation signals are fast changing and have zero mean. Therefore providing regulation services is ideal for data center batteries. However, since these signals are random in nature, and blindly following them may result sharp spikes in overall data center demand. Our strategy is to use battery in a complementary manner: use it to follow the regulation signals but shaving off the sharp peaks. Specifically, this paper makes the following contributions:
\begin{itemize}
\item We provide an optimization framework that enables batteries in data centers to be jointly optimized for both peaking shaving and regulation market participation. 
\item We demonstrate the optimal operation of batteries provide savings that are larger than the sum of the individually using batteries for each of the applications. Notably, we show that a data centers electricity bill can be reduced up to 30\%. 
\item Using real and synthetic data, we perform statistical analysis to show how batteries in data centers should be used to achieve the maximum saving. In particular, illustrate the relevant features of data center profiles that lead to the \emph{superlinear} gain.   
\end{itemize}

The paper is organized as follows. Section \ref{sec:background} provides some background on the topics studied in the paper and a review of the relevant literature. Section \ref{sec:formulation} gives the optimization formulation. Section \ref{sec:analysis} analyzes the solution of the optimization using synthetic data while in Section \ref{sec:eval} read data is used. Finally, Section \ref{sec:con} concludes the paper and outlines directions for future work.

\section{Background and Related Work}
\label{sec:background}

\begin{table}[ht]
	\centering
	\begin{tabular}{|l|l|}
		\hline
		Symbols & Definition\\
		\hline
		$\lambda_{elec}$ & \text{Energy charge  ($ \$/MWh$)} \\
		\hline
		$\lambda_{peak}$ & Peak demand charge ($\$/MW$)\\
		\hline
		$\lambda_c$ & Regulation capacity revenue ($\$/MWh$)\\
		\hline
		$\lambda_b$ & Battery degradation cost ($\$/MWh$)\\
		\hline
		$\lambda_{mis}$ & Regulation mismatch penalty ($\$/MWh$)\\
		\hline
		$r(t)$ & Normalized regulation signal\\
		\hline
		$b(t)$ & Battery power ($>0$ denotes discharge) ($MW$)\\
		\hline
		$s(t)$ & Data-center load ($MW$)\\
		\hline
		$y(t)$ & Energy baseline of data-center ($MW$)\\
		\hline
		$C$ &   Regulation capacity bid ($MW$)\\
		\hline
		$t_s$ & Time resolution (seconds)\\
		\hline
		$P$ & Battery power capacity ($MW$)\\
		\hline
		$E$ & Battery energy capacity ($MWh$)\\
		\hline
		$SoC_{ini}$& Initial battery energy percentile\\
		\hline
		$SoC_{min}$ & Minimal battery energy percentile\\
		\hline
		$SoC_{max}$ & Maximal battery energy percentile\\
		\hline
		$q$  & Superlinear saving ratio\\
		\hline
	\end{tabular}
	\caption{A summary of the notations used in this paper. }
	\label{Sec3:T1}
\end{table}

\subsection{Battery based Energy Storage Device}
\subsubsection{Battery energy storages in datacenter}
Batteries have become effective means in assisting end-consumer load managements and energy system operations~\cite{oudalov2006value}. Compared to fossil fuel based diesel generators, batteries are not constraint by locations and installation scales, and have instant response and ramping speed which making batteries ideal choices for responding to sudden events such as outage of the external electrical grid.

The operation of battery subjects to the power rating and the state of charge. The power rating is constraint by the type of battery technology and the battery management system. The state of charge describes the amount of energy stored in a battery, usually represented as a percentage with respect to the battery's energy capacity. The amount of energy a battery can charge into itself is constraint by the maximum safe level of the state of charge, while during the discharge process the state of charge must remain above a minimum level in order to avoid serious damage to battery cells~\cite{vetter2005ageing}.

In data centers, battery energy storages can be centrally installed in the electricity supply or distributively deployed to each server cabinet~(e.g. see \cite{WangEtAl2012} and the reference within). In either configuration, a data center is capable to seamlessly switch between its battery power supply and its grid power supply. The grid transmission and distribution system supplying power for the data center requires no modification since batteries are located and operated inside the data center, thus utilities typically view data centers as standard large commercial loads.

\subsubsection{Battery Cell Degradation Modeling}
A key factor in the operational planning of battery energy storages (BES) is its operating cost, a majority of which stems from the degradation of battery cells. In this study we target lithium-ion battery cells. Lithium-ion batteries have nonlinear degradation rate with respect to the cycle depth of discharge (DoD)~\cite{vetter2005ageing}. In irregular BES operations, the identification of cycles has been a germane issue. Although there are systematic cycle-counting algorithms such as the rainflow method that can be applied to an SoC profile for cycle identification~\cite{xu2014bess}, but these algorithms are too complicated to be applied into a convex optimization problem. 

Cycle-based battery degradation model is yet another well-adopted way to quantify the battery operating cost. In paper \cite{Guo2013}, the authors adopted an amortized cost to model the impact of per charging or discharging operation. Instead of accurately modeling the cost based on how fast/much/often a battery is charged or discharged, the authors assumed the same operating cost if the charging/discharging operation occurs and no cost when the battery is not used. This degradation model does not apply in our case. Because our battery is always employed either for peak shaving, providing regulation service or both, and the amount of electricity charged or discharged varies much, such a degradation model can not accurately capture the battery degradation cost.

In this paper, we consider a linear battery cost model based on the amount of electricity charged or discharged~\cite{Miguel2014} . In order to get the linearized battery degradation cost co-efficient, we normalize battery lifetime into the amount of energy a battery cell can process before reaching end-of-life using the relationship between the cycle depth and the number of cycles~\cite{ecker2014calendar}, and prorate the battery cell cost into a per-MWh cost with respect to the charged and discharged energy. This cost is a simplification of the complex chemical process that causes battery degradation. Also, we limit the operation of battery within certain SoC range in order to avoid the highly non-linear damage cost to battery cells.

\subsection{Peak Shaving}

Being a commercial demand, a data center must contract with a utility company for energy supply. In addition to the \emph{per-kWh} priced energy bill, utility companies also charge commercial consumers based on their peak demand because of its limited generation capacity and capital investment~\cite{alt2006energy}. The demand charge applies a \emph{per-kW} price to the consumer's maximum power draw during a billing cycle (typically a month).
The peak demand charge makes up a significant portion in the electricity bill. Therefore large consumers are incentivized to avoid sudden peaks in energy usage, an act commonly known as peaking shaving. With energy storage, a data center may shift some of its peak demand to low demand hours without interfering computation processes.

Authors in \cite{LiuWiermanChenEtAl2013,ZhouYaoGuanEtAl2015} analyzed how batteries can be used for peak shaving, authors in \cite{UrgaonkarEtAl2011,GovindanWangSivasubramaniamEtAl2013,ChenEtAl2008,PaulZhongBose2015, LiuLiuLowEtAl2014,XuLi2014,WangEtAl2015} have considered using batteries for demand response purposes, and \cite{ChenWangGiannakis2015} considered using batteries to perform arbitrage. Colocated data centers have been studied in \cite{AhmedIslamRen2015,ChenRenRenEtAl2015,TranDoRenEtAl2015,ZhangEtAl2015}.  In this paper we only consider batteries, but an overview of different types of storage devices and their placements in data centers for peak shaving can be found in \cite{WangEtAl2012}.

\subsection{Frequency Regulation}
Frequency regulation service involves the injection or withdrawal of active power from the power grid to maintain the system frequency. In deregulated electrical grids, power is allocated through double auction type of markets~\cite{KirschenEtAl2004}. These markets are cleared before the actual time of power delivery (usually one day and/or one hour prior to real-time). At real-time, to account for deviations from the forward markets, operators use a secondary market called the regulation market.  A resource procured in the regulation market follows the automatic generation control signal, which computes the area control error from frequency deviations and interchange power imbalances, and is paid by the grid operator in per-MW price with respect to the maximum active power that the resource is capable to inject or withdrawal~\cite{GloverEtAl2011}.

Data centers can utilize their batteries to participate in the regulation market. A data center reduces its instant electricity consumption by discharging batteries, and increases consumption by charging. Thus by incorporating batteries, a data center provides a responses to the regulation signal on top of the scheduled electricity consumption~\cite{XuDvorkinKirschenEtAl2016}.

Authors in \cite{ChenLiuCoskunEtAl2015, CaramanisPaschalidisCassandrasEtAl2012,LiuWiermanChenEtAl2013,ZhouYaoGuanEtAl2015,LiuEtAl2012,CamachoZhangChenEtAl2014,HeEtAl2012} considers how data centers can participate in various energy markets. In particular, \cite{ChenLiuCoskunEtAl2015} considers different types of storage devices and show that depending on their degradation costs, it may not be economical for batteries to participate in some markets. In contrast to existing research, we show that if regulation market participation and data center operation are considered jointly, using batteries can be profitable even if it is separately not economical for each of the two applications.

\section{Problem Formulation}
\label{sec:formulation}
In this section, we present the overall optimization framework in three steps. First, we consider using batteries only for peak shaving. Second, we consider using batteries only to provide regulation service. Finally, we formulate the optimization problem of using batteries for both applications. 
Before presenting the three steps, we first define the necessary notations (see Table \ref{Sec3:T1} for a complete list) and basic model assumptions. 



We consider a discrete time model, where time is discretized into steps of length $t_s$. The power consumption of the data center at time $t$ is denoted by $s(t)$. The granularity of $t_s$ equals 20 seconds, and the planning horizon T is one hour (T = 180). Over a time period T, the utility bill of a data center consists of both energy cost and peak power cost \cite{XuLi2014,GovindanWangSivasubramaniamEtAl2013}. Suppose the energy price is  $\lambda_{elec} (\$/MWh)$, then the total energy cost  is given by
\begin{equation}\label{Sec3:P1:1}
J^{elec} = \lambda_{elec} t_s \sum_{t=1}^{T} s(t)\,,
\end{equation} 

Suppose the peak power price is  $\lambda_{peak} (\$/Mw)$, then the peak demand charge is,
\begin{equation}
 J^{peak} = \lambda_{peak} \max_{1,\dots,T} s(t)\,,
\end{equation} 

Since the peak calculation is often at a much slower time-scale rather than every hour, we scale the $\lambda_{peak}$ accordingly and use an amortized hourly peak demand cost.

In the rest of the paper, the time constant $t_s$ is absorbed into the price coefficients for simplicity. Therefore, the total electricity bill of a data center over a time period of $T$ is 
\begin{equation}\label{Sec3:P1:3}
	J=\lambda_{elec} \sum_{t=1}^{T} s(t)+\lambda_{peak} \max_{1,\dots,T} s(t). 
\end{equation} 

Throughout this paper, we will make a key assumption, $s(t)$ is a deterministic and known signal. In essence, we assume that the future load of a data center can be predicted with perfect accuracy. This assumption is valid in this paper since we only consider the aggregate load or a data center, and that load exhibits relatively stable daily pattern \cite{WangEtIISWC2013}. The load becomes less predictable at more granular time scale, and understanding load profiles at different levels in a data center is an active direction of future research.

\subsection{Peak shaving}
Peak shaving reduces the total electricity bill of a data center by minimizing the second term in \eqref{Sec3:P1:3}, while taking the properties and degradation of batteries into account. Let $b(t)$ denote the power injected by the battery into the data center, thus $s(t)-b(t)$ represents the power consumed from the power grid. The optimization problem is
\begin{subequations}\label{Sec3:P2:1}
\begin{align}
 \min_{b(t)} \;
 &  \lambda_{elec} \sum_{t=1}^{T} {[s(t)-b(t)]} + \lambda_{peak} \underset{t=1...T}{\text{max}}{[s(t)-b(t)]} \\
 &  + \lambda_b \sum_{t=1}^{T} {|b(t)|} \nonumber \\ 
 \text{s.t.}  &  -P \leq b(t) \leq P \label{eqn:pt} \\
 &   SoC_{min} \leq \frac{SoC_{ini} \* E + \sum_{t=1}^{t}b(\tau) \* t_s}{E} \leq SoC_{max}  \label{eqn:soc}
\end{align}
\end{subequations} 
The constraints in \eqref{eqn:pt} and \eqref{eqn:soc} represent the power limit of the battery and the state of charge limit of the battery, respectively.

The last summation term in the objective function models the battery operation cost, i.e., the degradation cost, where $\lambda_d$ is the per-MWh cost.  We denote that the optimal battery scheme solved by the above optimization problem is $b^{p}(t)$ and the minimum electricity bill is $J^{p}$.

\subsection{Regulation service}
In this section, we model batteries in the data center as a resource to participate in the regulation market ~\cite{PJM_DR}. We formulate a revenue maximization problem based on simplified regulation market policies~\cite{XuDvorkinKirschenEtAl2016}. In the regulation market, the grid operator pays a per-MW option fee ($\lambda_c$) to a resource for the stand-by regulation power capacity ($C$) that the resource can provide for the grid. While during the regulation procurement period, the resource is subjected to a per-MWh regulation mismatch penalty ($\lambda_{mis}$) for the absolute error between the instructed dispatch and the resource's actual response. 

Let's denote $r(t)$ as the normalized regulation signal send out by the system operator. The regulation revenue ($R$) is,
\begin{equation}\label{Sec3:P3:1}
	R=\lambda_c C - \lambda_{mis} \sum_{t=1}^{T}{|b(t)-Cr(t)|} - \lambda_b \sum_{t=1}^{T} |b(t)|,
\end{equation} 
where we model the battery operating cost as reductions in the regulation revenue.

As a demand response resource, the data center forecasts its load curve and sends to the grid operator. During the regulation, the grid operator subtracts the data center's reported load curve from the real-time measured data center power consumption to calculate the data center's regulation response. By assuming data center can perfectly forecast its demand, we can isolate the regulation from the energy usage and directly simplify the data center's regulation response as the response of the batteries.
Therefore the revenue maximization problem subjects to both battery operation constraints and non-zero regulation capacity: 
\begin{subequations}\label{Sec3:P3:2}
	\begin{align}
	\max_{C, b(t)} \;
	&  \lambda_c C - \lambda_{mis} \sum_{t=1}^{T}{|b(t)-Cr(t)|} - \lambda_b \sum_{t=1}^{T} |b(t)| \\
	\text{s.t.}  &   \ C \geq 0 \label{eqn:pt1} \\
	&  -P \leq b(t) \leq P \label{eqn:pt2} \\
	&   SoC_{min} \leq \frac{SoC_{ini} \* E + \sum_{t=1}^{t}b(\tau) \* t_s}{E} \leq SoC_{max}  \label{eqn:soc2}
	\end{align}
\end{subequations} 

The constraints in (\ref{eqn:pt2}) and (\ref{eqn:soc2}) are the same as the peak shaving case, denote the power and SoC limit of battery. The first constraint \ref{eqn:pt1} guarantees the regulation capacity should be non-negative. In other works, a participant in the regulation market should have a non-zero regulation power capacity.

In the proposed regulation revenue maximization problem, we do not consider the effect of providing regulation service on the data center electricity bills. Recall that regulation is a service managed by grid operators, while as a end-consumer the data center's electricity supply contract with the utility is unchanged, thus the data center still subjects to the energy and peak demand charge, and the overall revenue $J^r$ is
\begin{equation}\label{Sec3:P2:3}
	J^{r} = \lambda_{elec} \sum_{t=1}^{T} {[s(t)-b^{r}(t)]} + \lambda_{peak} \underset{t=1...T}{\text{max}}{[s(t)-b^{r}(t)]}-{R^{*}}\,,
\end{equation}

\begin{figure}
	\centering
	\includegraphics[width= 0.6 \columnwidth, height= 0.5 \columnwidth]{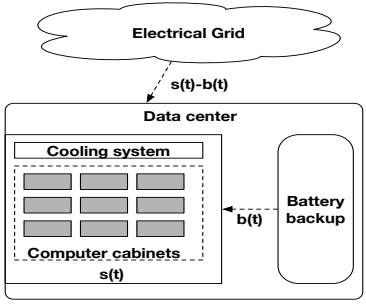}
	\caption{Main components of a typical data center. Notations for power demand s(t), battery output b(t) and grid consumption s(t)-b(t).}
	\label{fig3_2_1}
\end{figure}

\subsection{Both regulation service and peak shaving}

In this work we offer a pioneering view by considering regulation and peak shaving together in a single decision problem. The combined problem minimizes the total charge in electricity including regulation revenue: 
\begin{subequations}\label{Sec3:P4:1}
	\begin{align}
		\min_{C, b(t), y(t)} \;
		&  \lambda_{elec} \sum_{t=1}^{T} {[s(t)-b(t)]} + \lambda_{peak} \underset{t=1...T}{\text{max}}{[s(t)-b(t)]} \\
		&  + \lambda_b \sum_{t=1}^{T} {|b(t)|} \nonumber \\ 
		&  - (\lambda_c \* C - \lambda_{mis} \sum_{t=1}^{T}{|-s(t)+b(t)+y(t)-Cr(t)|})\nonumber\\
		\text{s.t.}  &  C \geq 0 \label{co:pc}\\
		& y(t) \geq 0 \label{co:py}\\
		&  -P \leq b(t) \leq P \label{co:pt} \\
		&   SoC_{min} \leq \frac{SoC_{ini} \* E + \sum_{t=1}^{t}b(\tau) \* t_s}{E} \leq SoC_{max}  \label{co:soc}
	\end{align}
\end{subequations} 

Similar to the former two optimization problems, we have the zero net energy change constraint, the battery power and SoC constraints. What's more, (\ref{co:pc}) and (\ref{co:py}) defines the constraints for a data center to participate in the regulation. (\ref{co:pc}) guarantees that the data center has a non-negative regulation capacity. And $y(t)$ is the load curve forecast the data center submits to the grid operator for regulation response evaluation. Since we are considering both services at the same time, it is crucial to include the forecast load curve in the formulation to model the actual penalty of regulation mismatch.     


We denote the optimal battery scheme by solving the optimization problem in (\ref{Sec3:P4:1}) as $b^{*}(t)$ and the minimum electricity bill achieved as $J^{*}$. 

\subsection{Superlinear gain}
Our result highlight that the electricity bill saving from implementing the joint-optimization could exceed the sum from both sides, which could be expressed as the following mathematical form,
\begin{equation}\label{Sec3:P5:1}
J-J^{*} > (J-J^{r})+(J-J^{p})\,,
\end{equation} 

\begin{figure}
	\centering
	\begin{subfigure}[b]{0.475\columnwidth}
		\centering
		\includegraphics[width=\columnwidth, height = \columnwidth]{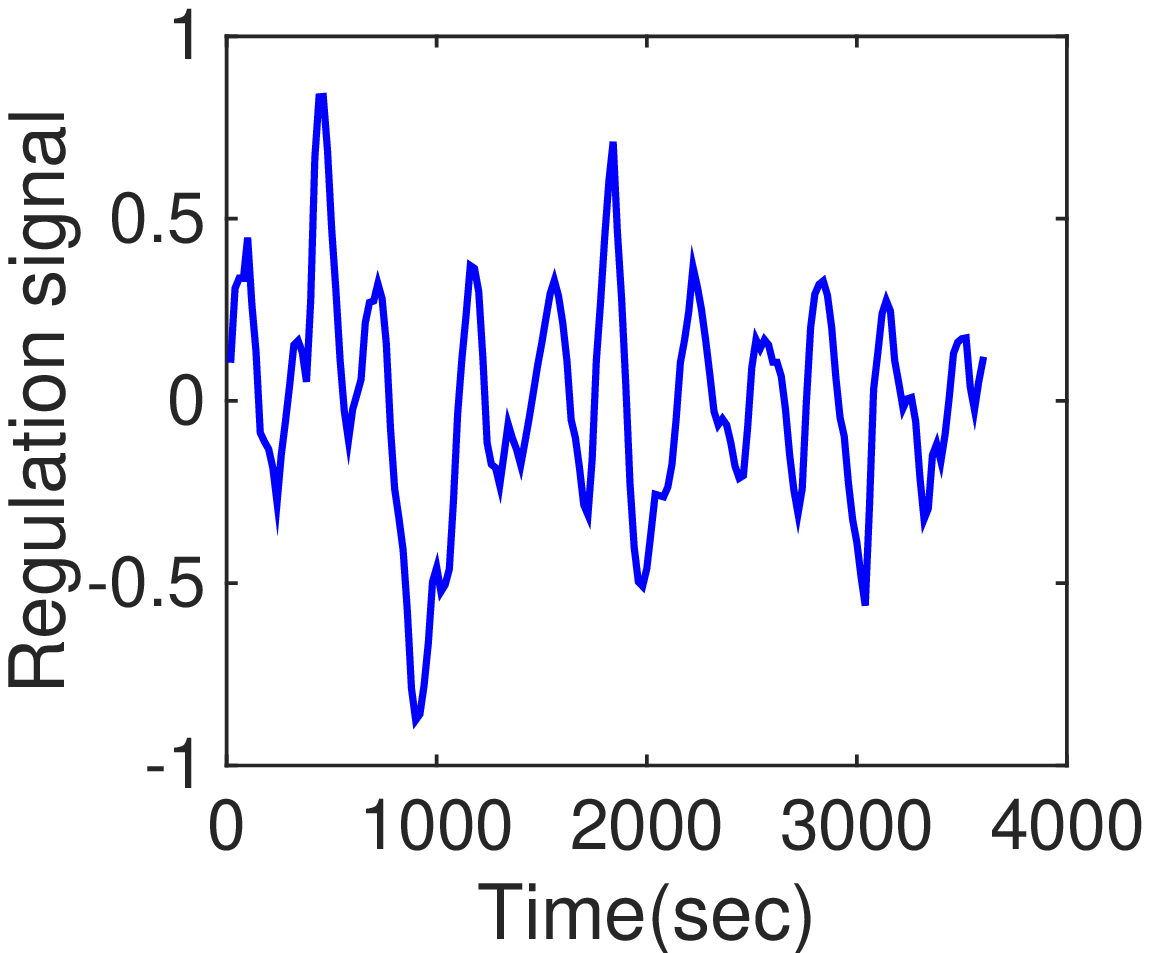}
		\caption[Network2]%
		{{\small One hour regulation signal}}    
		\label{fig3_5_1}
	\end{subfigure}
	\hfill
	\begin{subfigure}[b]{0.475\columnwidth}  
		\centering 
		\includegraphics[width=\columnwidth, height = \columnwidth]{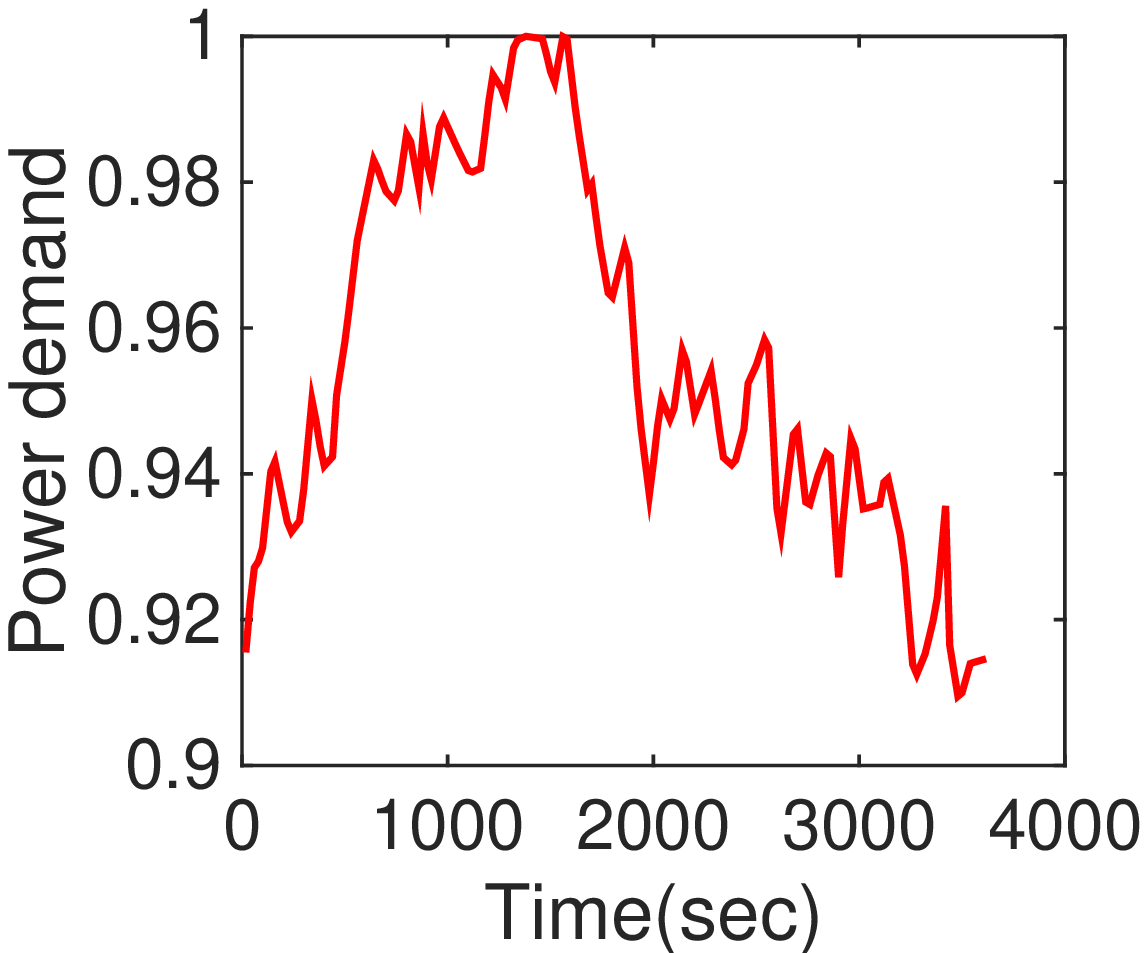}
		\caption[]%
		{{\small One hour power demand}}    
		\label{fig3_5_2}
	\end{subfigure}
	\caption[ The average and standard deviation of critical parameters ]
	{\small One-hour regulation signal and power demand} 
	\label{fig3_5_3}
\end{figure}

Table \ref{Sec3:T2} shows the electricity bills of a data-center within one hour under the four scenarios: the unoptimized bill (batteries are left idle), using battery only for regulation service, using battery only for peak shaving, and using battery for both services. Fig.\ref{fig3_5_3} gives the regulation signal and power demand signal for that specific hour. The data center power demand traces used in this paper comes from Microsoft over a six-month period. The power values are normalized to their peak demand.

Note within this papar, we assume that the datacenter power demand and regulation signal are known and solve the optimization problem offline. But the offline result should not be too far from the online one, since the online method does not rely on complete knowledge of the future power demand patterns and regualtion signals. For online implementation, a good bidding capacity could be choosen using historical data. And it is not hard to show that under linear battery cost model, a simple control scheme in which the battery exactly follows the regulation signal unless hitting its power and SoC limits, is optimal. Therefore, our conclusion on the potential superlinear gain from the offline optimization also apply to online implementation. 

\begin{table*}
	\centering
	\begin{tabular}{|l|l|l|l|l|l|l|}
		\hline
		& Total bill & Bill saving & Energy charge & Peak charge & Battery cost & Regulation rev\\
		\hline
		Original 			& 73.81 & \todo{0}       & 44.92 & 28.89  &0 & 0\\ 
		\hline	
		Regulation only		  & 59.66 & \todo{14.15}& 44.92 & 54.55 & 25.90 & 65.71\\
		\hline
		Peak shaving 	& 73.53 & \todo{0.28}  &  44.92 & 28.35 & 0.26 & 0\\
		\hline
		Joint-optimization			     & 52.46 & \todo{21.35}    & 44.92 & 42.86 & 19.48 & 54.80\\
		\hline
	\end{tabular}
	\caption{Electricity bills}
	\label{Sec3:T2}
\end{table*}

\section{Analysis of the super-linear gain}
\label{sec:analysis}
In this section, we aim to provide some insights of the super-linear saving gained by the joint-optimization method. One natural question is, when we shall have the super-linear saving? In order to answer this question, we need a better abstraction to capture the power demand features.

We enhance the abstraction model proposed by \cite{WangEtIISWC2013}. For a \emph{single} peak demand, we identify three important attributes, namely the peak height, peak width and peak shape and then investigate how these attributes influence the effectiveness of joint-optimization method. Besides a single peak, we also care about the correlation between peaks. We define the \emph{Number of Contiguous Peaks (NOCP)} as a key attribute for contiguous peaks and show its effect in the super-linear gain.

\subsection{Peak abstraction}
\begin{figure}[H]
	\centering
	\includegraphics[width= 0.95 \columnwidth, height = 0.618 \columnwidth]{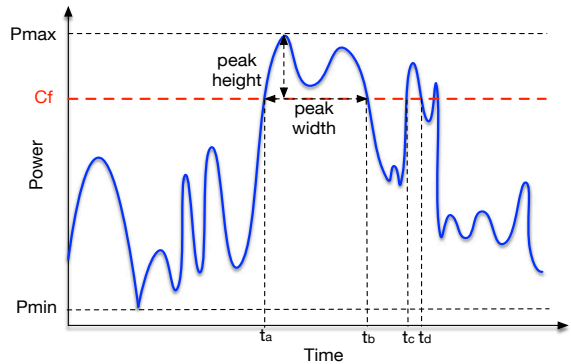}
	\caption{Notations for peak abstraction}
	\label{fig4_1_1}
\end{figure}

Given a power demand series, let $p_{min}$ and $p_{max}$ denote the minimum and maximum power demand over the demand series. Then the variation range of power consumption, represented by $d$, could be calculated as $d = p_{max} - p_{min}$.
In order to differentiate between power peaks and power valleys, we define $C_f$ as the threshold for peaks. If the power demand within a time interval is above $C_f$, we call it a peak. Otherwise, if the power demand is below $C_f$, it is a valley. Here, we determine the threshold in terms of the $f$, the percentage of dynamic range. Thus, the peak threshold $C_f$ is calculated using equation (\ref{Sec4:P1:2}).

\begin{equation}\label{Sec4:P1:2}
C_f = (1-f)  d + p_{min}\,,
\end{equation}
Once we set a specific threshold $C_f$, the demand series are divided into a sequence of peaks and valleys. In the scope of this paper, we only consider peaks. For example, the power demand is above $C_f$ during time interval $[t_a, t_b]$ and $[t_c, t_d]$ in Fig.\ref{fig4_1_1}, thus they are two typical examples of peaks.

For a single peak, we use the following attributes for peak abstraction. Assume a peak starts from time $t_a$ and ends at $t_b$, and we define the peak height, peak width, peak shape as following:
\begin{itemize}
	\item Peak height(PH):
	$\frac{max_{t_a \leq t \leq t_b}{[s(t)-C_f]}}{d}$
	\item Peak width(PW): $t_{b}-t_a$
	\item Peak shape: peak shapes are characterized into two categories: triangular-like shape and rectangular-like shape
\end{itemize}

For multiple peaks, we care about their correlations. For instance, is there a high probability that two or more peaks happen contiguously? What's the average number of adjacent peaks? Therefore, we define \emph{Number of Contiguous Peaks (NOCP)} as a key attribute to depict the correlation between peaks. If the time interval between the end time of a prior peak and the start time of the following peak is within a threshold (eg. 2 minutes), we count it as a continuous peak pair. The number of contiguous peaks will continue increase until we the time interval between two peaks exceeds the threshold.

\subsection{Single peak characteristics vs. superlinear gain }
As defined in the previous section, there are three attributes for a single peak: peak height, peak width and peak shape. Therefore, we could characterize the peaks into 8 categories as shown in Table \ref{Sec4:T1}.
\begin{table}[H]
	\centering
\begin{tabular}{ |l|l|l| }
	\hline
	\multicolumn{3}{|c|}{Rectangular/ Triangular peaks} \\
	\hline
	\backslashbox{width}{height} & low  & high \\
	\hline
	narrow & narrow and low & narrow and high \\
	\hline
	wide & wide and low & wide and high \\
	\hline
\end{tabular}
\caption{Rectangular/Triangular peak characterization}
\label{Sec4:T1}
\end{table}

Corresponding to the 8 categories of peaks in Table \ref{Sec4:T1}, we design 8 kinds of synthetic demand traces. For each trace, we assume that the base-load is 1 MW, and the peak duration and peak power are set as below.
\begin{itemize}
	\item \textbf{Narrow and low:}\\ Peak duration = 120s, peak demand = 1.33MW.
	\item \textbf{Narrow and high:} \\ Peak duration = 120s, peak demand = 2MW.
	\item \textbf{Wide and low:} \\Peak duration = 600s, peak demand = 1.33MW.
	\item \textbf{Wide and high:}\\ Peak duration = 600s,  peak demand = 2MW.
\end{itemize}

Considering rectangular-shape and triangular-shape peaks, there are 8 kind of synthetic traces in total. For each category, we simulate for 100 times in random days and hours under four scenarios: not using the battery at all, using battery only for regulation service, using battery only for peak shaving and the joint-optimization method.  Denote the electrical bills under the four scenarios as $J$, $J^{r}$, $J^{p}$, $J^{*}$, and if $J-J^{*} > (J-J^{r})+(J-J^{p})$ holds, we say that we obtain the super-linear gain by the proposed joint-optimization method. The statistical probability of obtaining the super-linear gain under eight kinds of peaks is listed in Table \ref{Sec4:T2}.
\begin{table}[H]
	\centering
	\begin{tabular}[width= 0.95 \columnwidth]{|l|l|l|l|l|}
		\hline
		& \multicolumn{2}{c|}{Rectangular} & \multicolumn{2}{c|}{Triangular} \\
		\hline
		\backslashbox{peak width}{peak height} & low  & high  & low  & high \\
		\hline
		narrow & 0.58 & \todo{0.1} & 0.56 & \todo{0.07} \\
		\hline
		wide & \todo{0.93} & 0.9 & \todo{0.75} & 0.24\\
		\hline
	\end{tabular}
	\caption{Probability of obtaining the super-linear gain under different category of peaks}
	\label{Sec4:T2}
\end{table}
The results given in Table \ref{Sec4:T2} highlight a number of important observations about the correlations between peak attributes and obtaining the super-linear gain.

First, note that when the peak is \emph{sharp} (narrow and high), the probability of wining the nonlinear gain is very low in both the rectangular and triangular cases, $0.1$ and $0.07$ respectively. However, when the peak is \emph{flat} (wide and low), the probability of obtaining the super-linear gain is high. In order to explore the main difference between the sharp peak case and flat peak case, we plot the grid power consumption and battery SoC curve under regulation only, peak shaving only as well as the joint-optimization strategy for detailed analysis.
\begin{figure}[ht]
	\centering
	\begin{subfigure}[b]{0.95\columnwidth}
		\centering
		\includegraphics[width=\columnwidth, height= 0.35 \columnwidth]{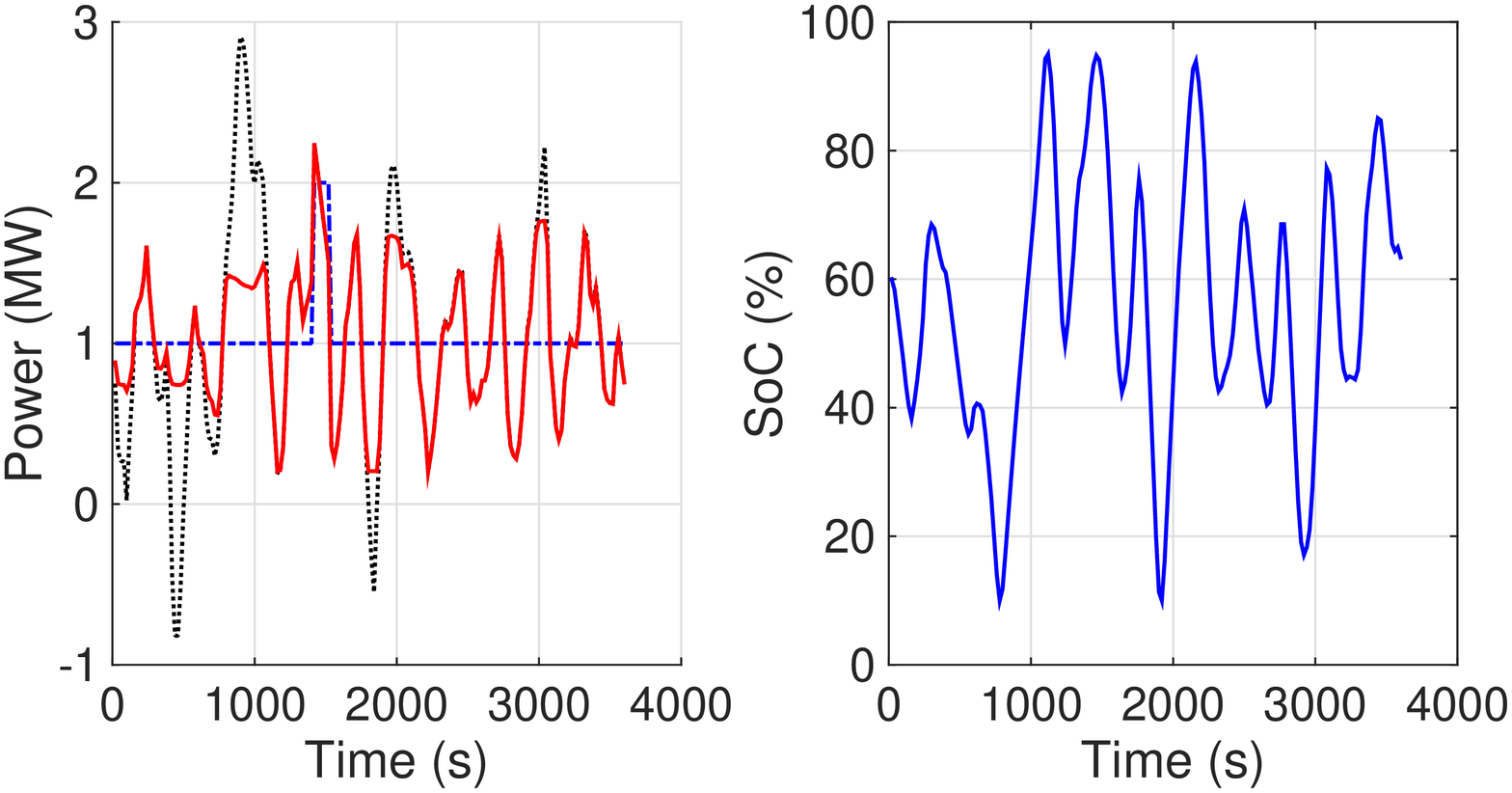}
		\caption[Network2]%
		{{\small Regulation only}}
		\label{Sec4:P2:1}
	\end{subfigure}
	\hfill
	\begin{subfigure}[b]{0.95\columnwidth}
		\centering
		\includegraphics[width=\columnwidth, height= 0.35 \columnwidth]{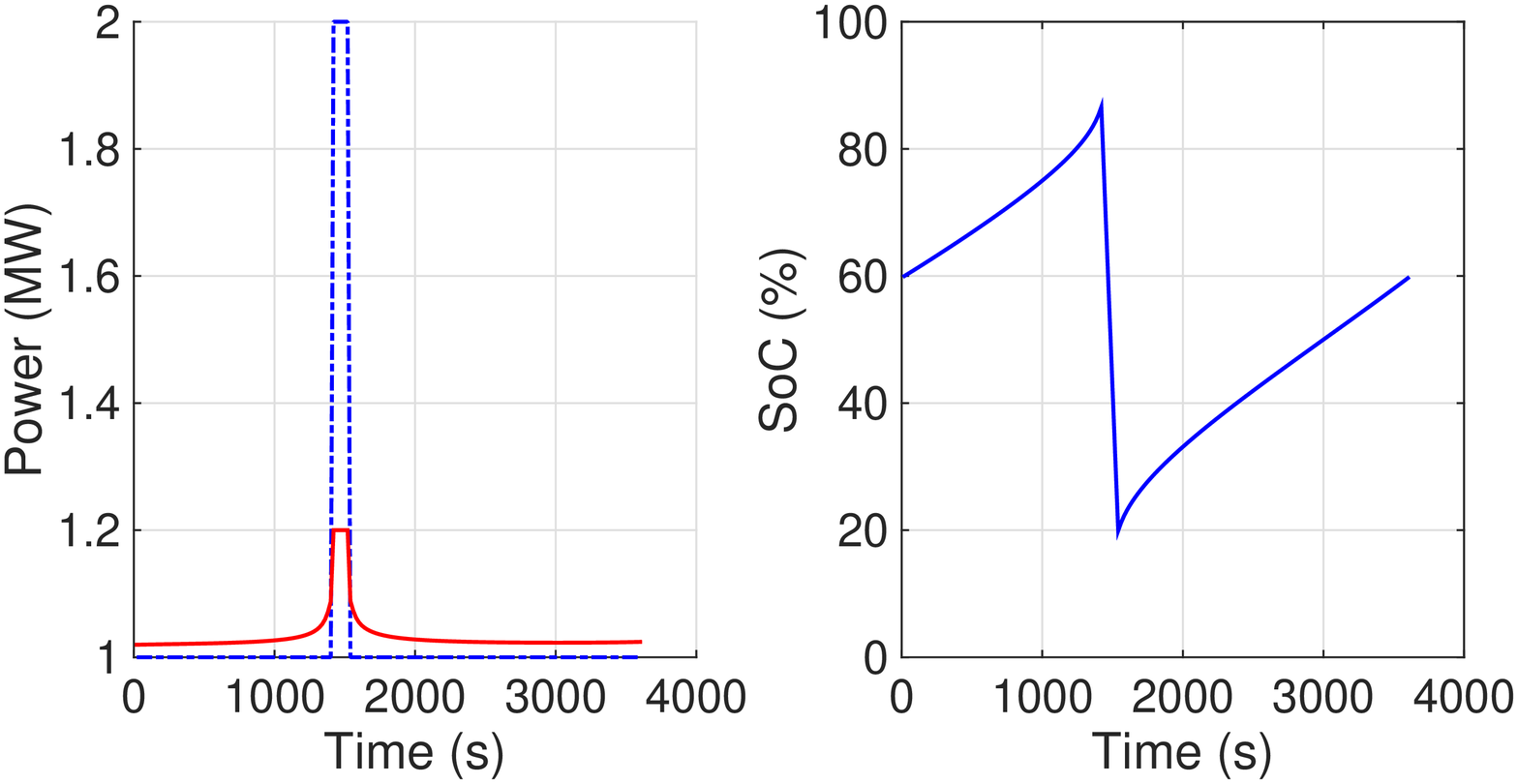}
		\caption[]%
		{{\small Peak shaving only}}
		\label{Sec4:P2:2}
	\end{subfigure}
	\vskip\baselineskip
	\begin{subfigure}[b]{0.95\columnwidth}
		\centering
		\includegraphics[width=\columnwidth, height= 0.35 \columnwidth]{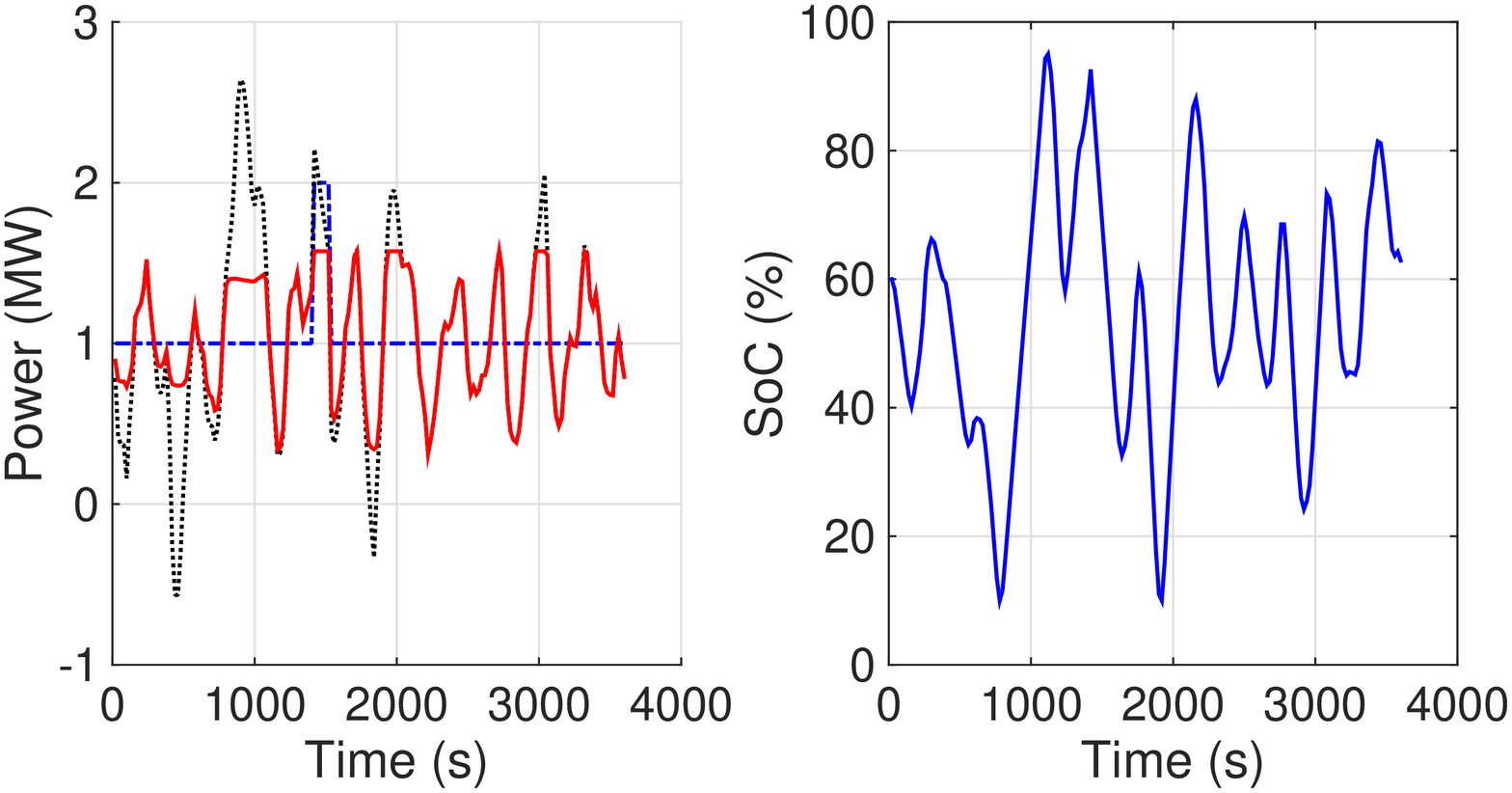}
		\caption[]%
		{{\small Joint optimization}}
		\label{Sec4:P2:3}
	\end{subfigure}
	\quad
	\begin{subfigure}[b]{0.95\columnwidth}
		\centering
		\includegraphics[width=\columnwidth, height= 0.35 \columnwidth]{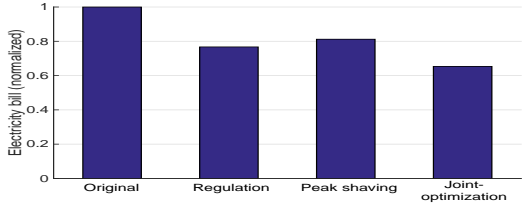}
		\caption[]%
		{{\small Electricity bills comparison}}
		\label{Sec4:P2:4}
	\end{subfigure}
	\caption[ The average and standard deviation of critical parameters ]
	{\small Electricity bills in sharp peak. Notion: For figure a, b, c, in the left plots, the blue dash-dotted line denotes the original demand s(t); the black dotted line denotes s(t)+Cr(r); and the red line denotes the grid demand s(t)-b(t). Battery SoC curve is the right half.}
	\label{Sec4:P2:s1}
\end{figure}

\begin{figure}[ht]
	\centering
	\begin{subfigure}[b]{0.95 \columnwidth}
		\centering
		\includegraphics[width=\columnwidth, height= 0.35 \columnwidth]{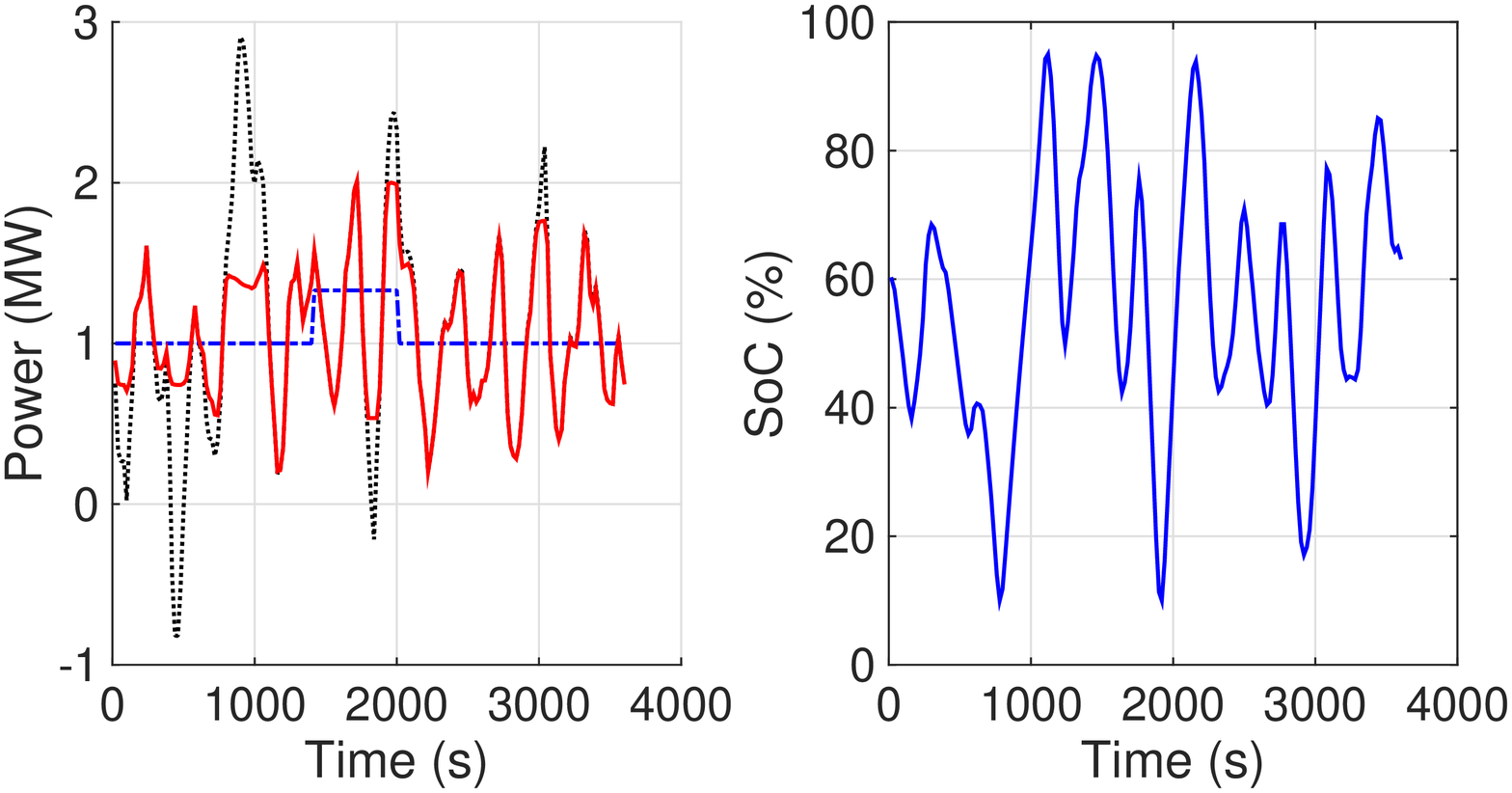}
		\caption[Network2]%
		{{\small Regulation only}}
		\label{Sec4:P2:5}
	\end{subfigure}
	\hfill
	\begin{subfigure}[b]{0.95 \columnwidth}
		\centering
		\includegraphics[width=\columnwidth, height= 0.35 \columnwidth]{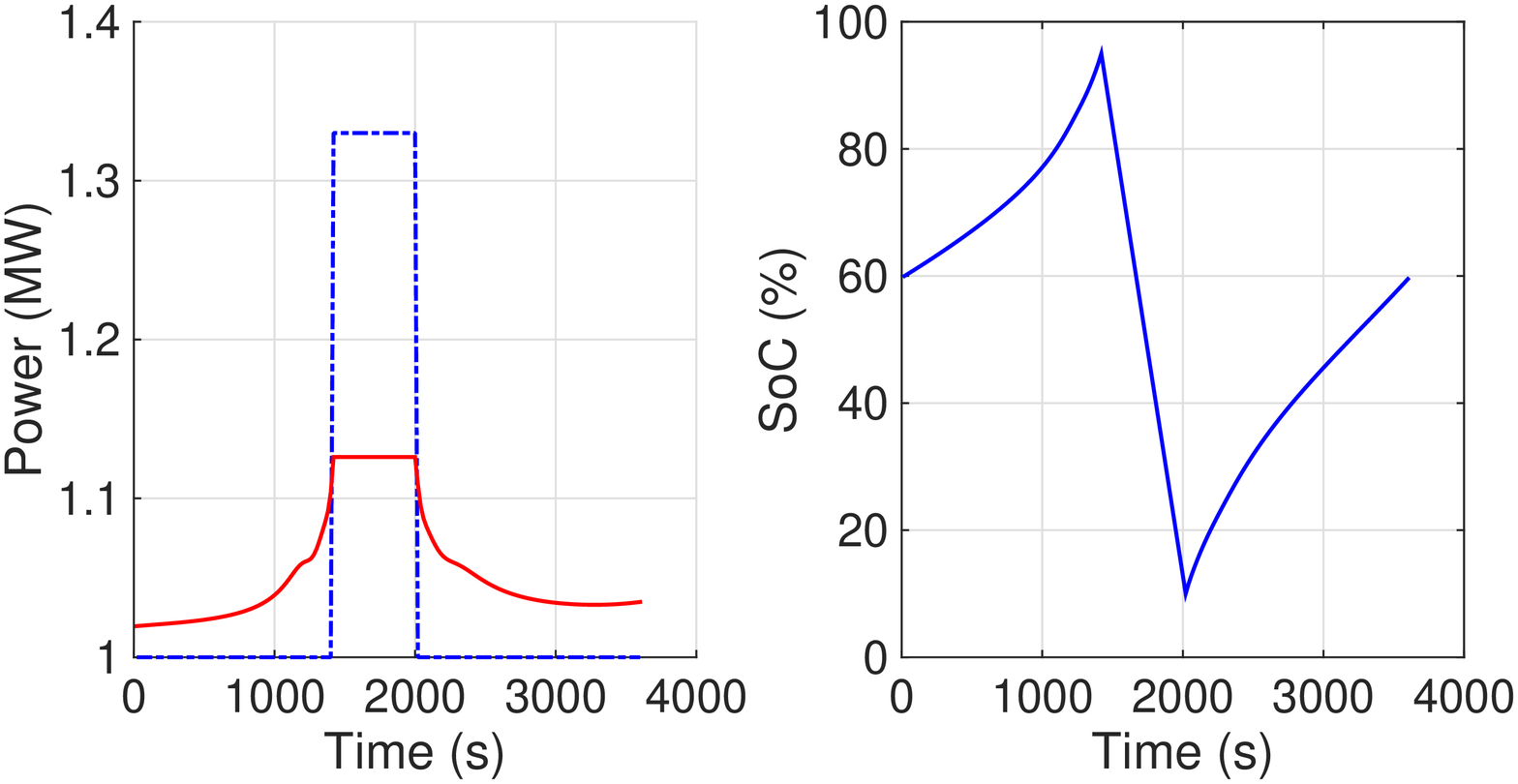}
		\caption[]%
		{{\small Peak shaving only}}
		\label{Sec4:P2:6}
	\end{subfigure}
	\vskip\baselineskip
	\begin{subfigure}[b]{0.95 \columnwidth}
		\centering
		\includegraphics[width=\columnwidth, height= 0.35 \columnwidth]{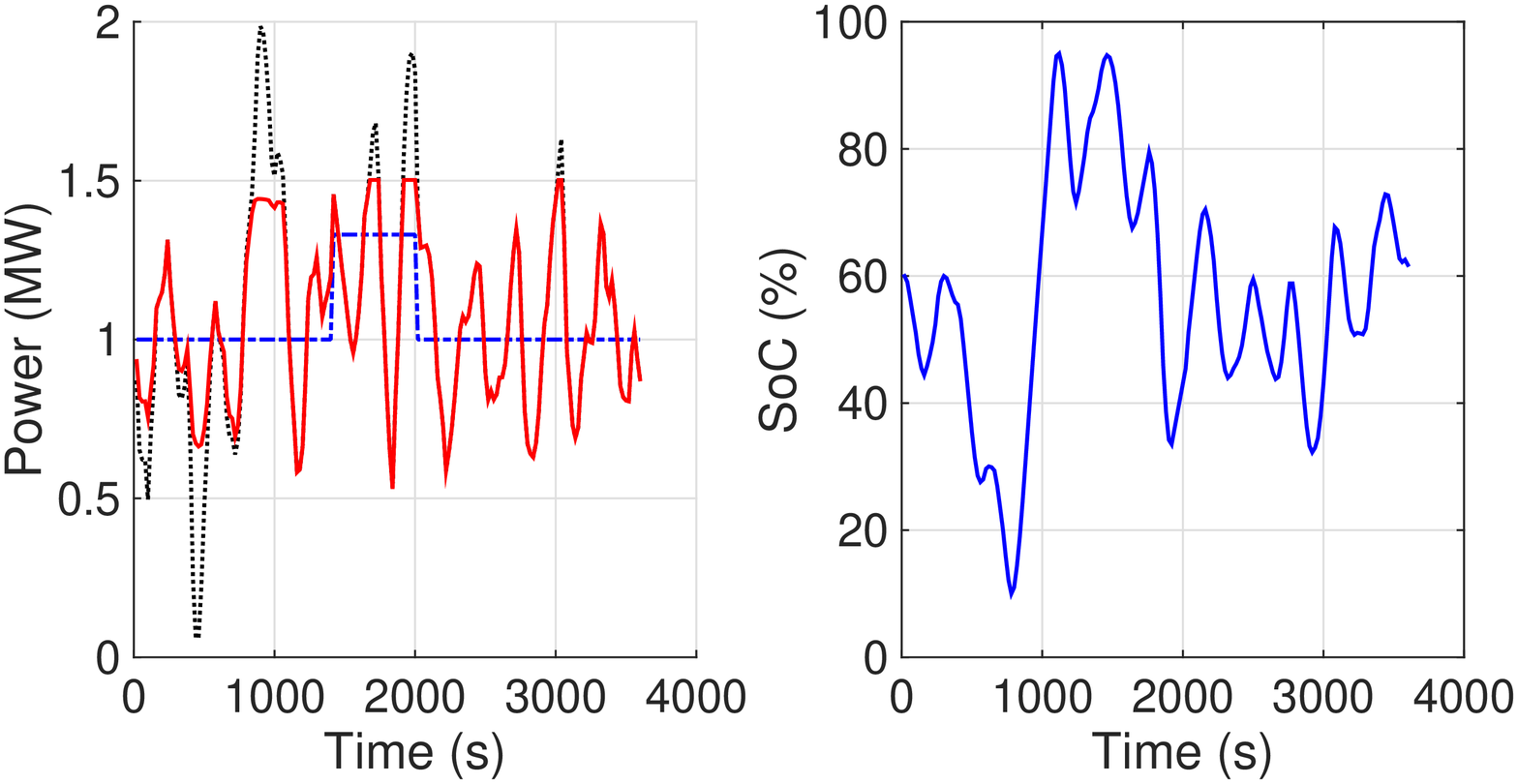}
		\caption[]%
		{{\small Joint optimization}}
		\label{Sec4:P2:7}
	\end{subfigure}
	\quad
	\begin{subfigure}[b]{0.95 \columnwidth}
		\centering
		\includegraphics[width=\columnwidth, height= 0.35 \columnwidth]{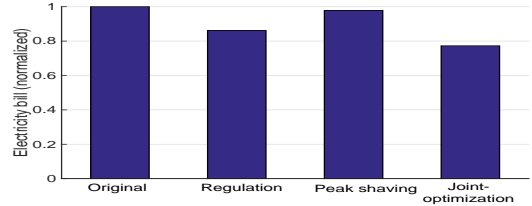}
		\caption[]%
		{{\small Electricity bills comparison}}
		\label{Sec4:P2:8}
	\end{subfigure}
	\caption[ The average and standard deviation of critical parameters ]
	{\small Electricity bills in flat peak. Notion: For figure a, b, c, in the left plot, the blue dash-dotted line denotes the original demand s(t); the black dotted line denotes s(t)+Cr(r); and the red line denotes the grid demand s(t)-b(t). Battery SoC curve is the right half.}
	\label{Sec4:P2:s2}
\end{figure}

By comparing Fig. \ref{Sec4:P2:s1} and Fig. \ref{Sec4:P2:s2}, we find that the main difference lies in the peak shaving part. For a sharp peak, the battery could shave it completely without hiting the SoC bound as suggested in Fig.\ref{Sec4:P2:s1} (b). Thus, in sharp peak case, we could save a lot from doing peak shaving only and won't obtain the super-linear saving. However, when the peak is flat, it takes a lot more battery energy to shave the same height of peak. As seen from Fig.\ref{Sec4:P2:s2} (b), the battery actually hits its SoC bound. In such condition, the extra cost of employing battery to do peak shaving gets close to the saving from reduced peak charge. This argument is verified by Fig.\ref{Sec4:P2:s2} (d), where we can see doing peak shaving only does not reduce the bill much. But if we consider to take part in the regulation market at the same time, the randomness of regulation signal helps break down the one flat peak into several sharp peaks, and we could save more from doing peak shaving on top of providing regulation service. And this is where the super-linear saving comes from.

Another interesting observation is that for the \textquotedblleft wide and high\textquotedblright\ peak case, while the probability of getting the super-linear gain for rectangular shape peaks is 90\%, the probability for triangular shape peaks is only 24\%. This again share the same reasons as our above analysis, since for triangular peaks, by nature, has some kind of \textquotedblleft sharpness\textquotedblright. As an example, consider a wide and high magnitude triangular peak, the battery could use much less battery energy to shave the same height of peak compared to shave a rectangular peak with the same width. So at many times, we may save a lot from doing peak shaving only and won't obtain the super-linear saving.

A final observation is that for a single \emph{small} area peak, which is both narrow and low, the probability of obtaining the super-linear gain and the probability of not obtaining it is roughly even. It is not surprising to have such result since a narrow and low peak can be considered as an \textquotedblleft immediate state\textquotedblright\  between sharp and flat peaks. Since it is narrow, thus the battery could shave the peak completely without much energy consumption.  However, as it is a low peak, there is not much room to have further savings.

\begin{figure}[ht]
	\centering
	\begin{subfigure}[b]{\columnwidth}
		\centering
		\includegraphics[width=\columnwidth, height= 0.35\columnwidth]{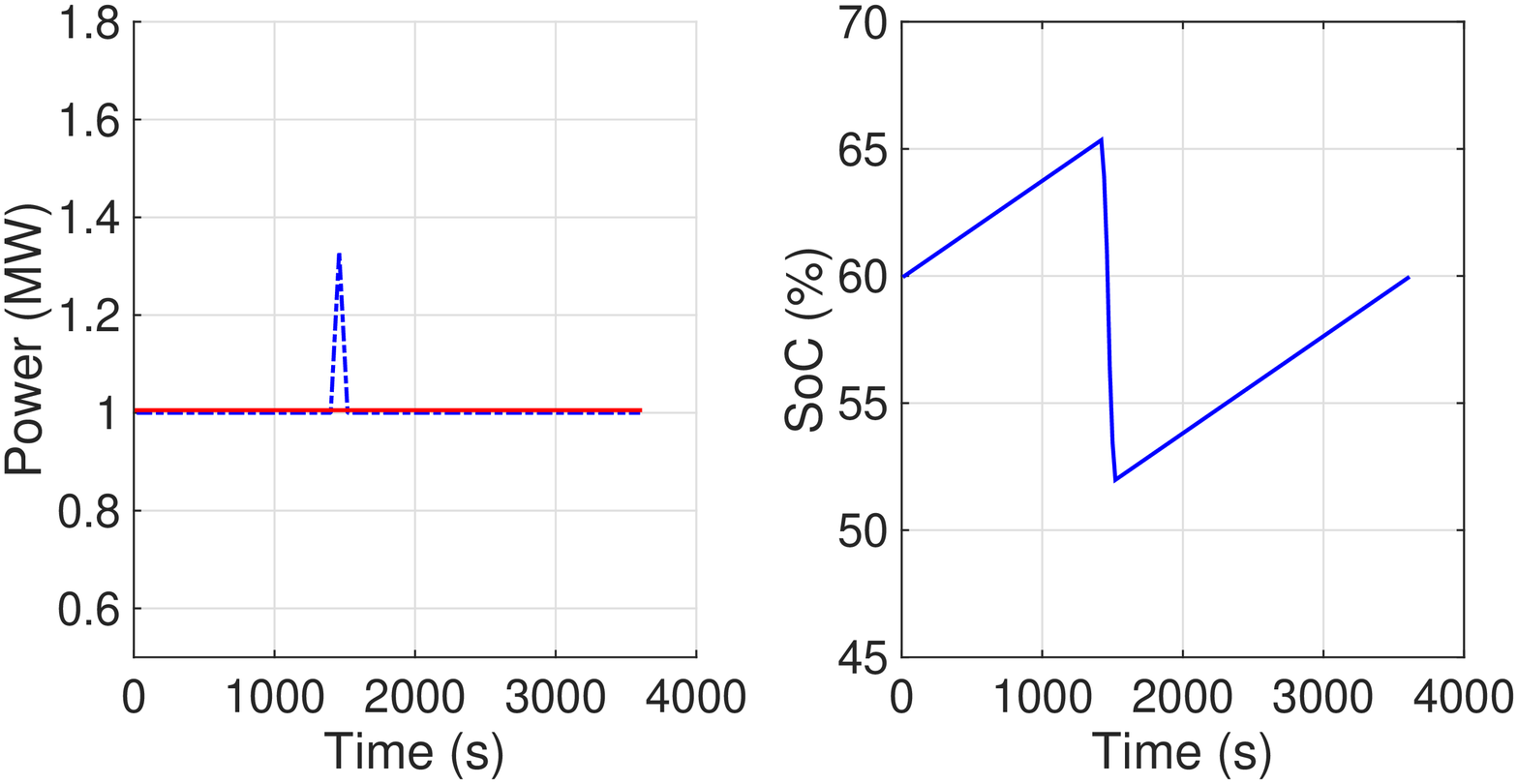}
		\caption[Network2]%
		{{\small Peak shaving only}}
		\label{Sec4:P2:9}
	\end{subfigure}
	\hfill
	\begin{subfigure}[b]{\columnwidth}
		\centering
		\includegraphics[width=\columnwidth, height= 0.35\columnwidth]{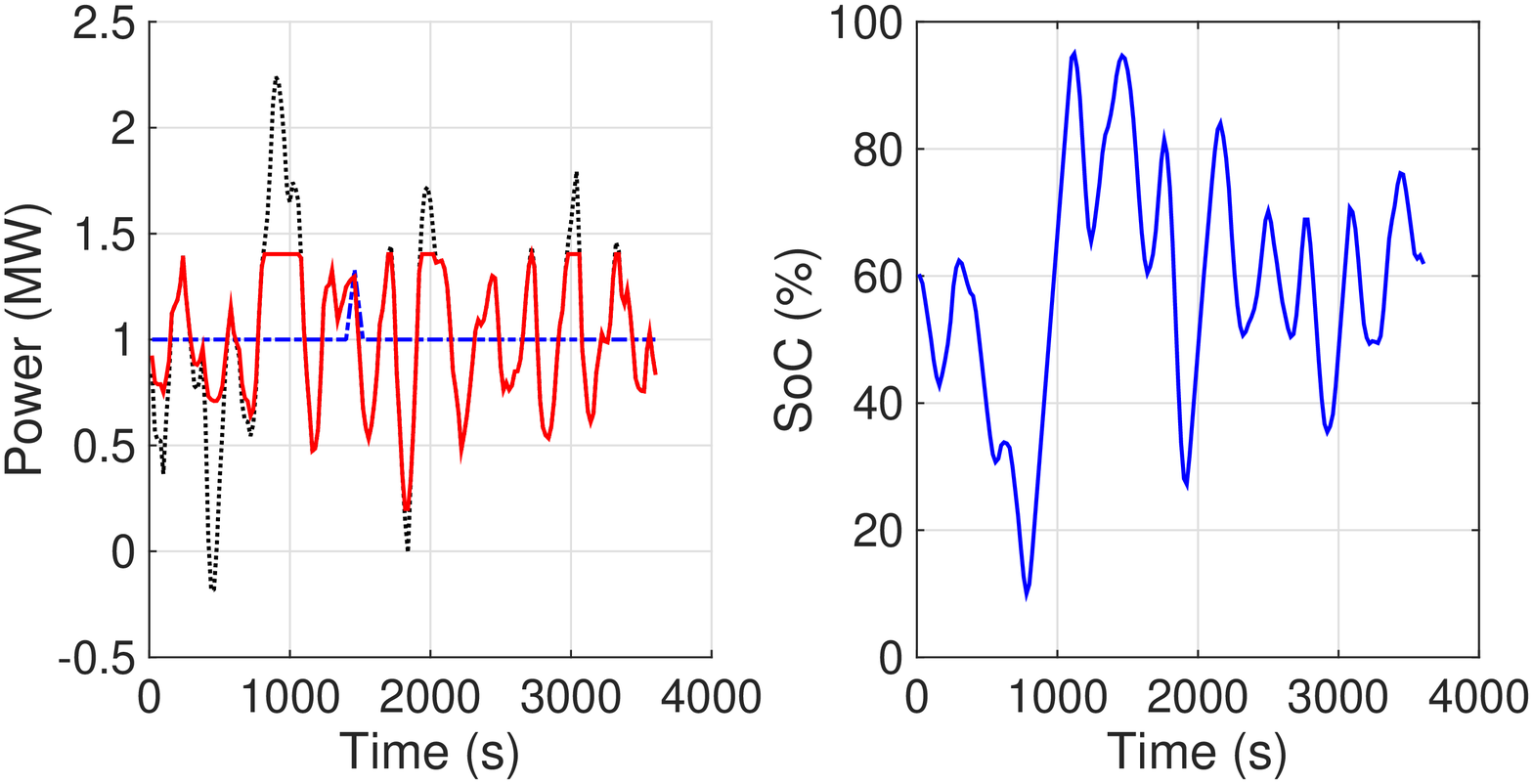}
		\caption[]%
		{{\small Joint optimization}}
		\label{Sec4:P2:10}
	\end{subfigure}
	\caption[ Low and narrow peak]
	{\small Grid consumption and battery SoC under narrow and low peak case. Notion of lines are the same as Fig. \ref{Sec4:P2:s1} and \ref{Sec4:P2:s2}}
	\label{Sec4:P2:s3}
\end{figure}

\subsection{Contiguous peaks characterization vs. super-linear gain}
In section 4.2, we discussed an interesting observation that when a single peak is both low and narrow, the probability of obtaining the super-linear or not obtaining is roughly even. This holds for both rectangle shaped peaks and triangle shaped single peaks. However, if multiple this type of \textquotedblleft low and narrow\textquotedblright\ peaks occur in a sequence, will it increase, decrease or not affect the probability of gaining the super-linear gain?

\begin{figure}[ht]
	\centering
	\includegraphics[width= \columnwidth, height = 0.618 \columnwidth]{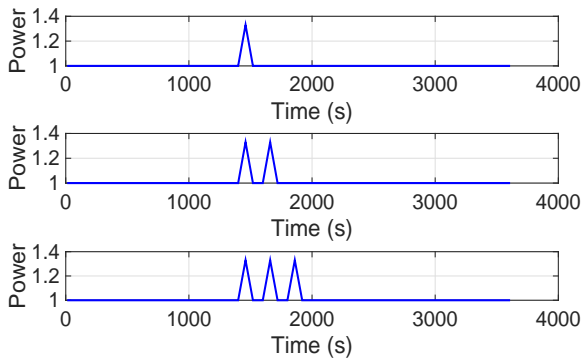}
	\caption{Consecutive small triangle peaks}
	\label{Sec4:P3:1}
\end{figure}

Figure \ref{Sec4:P3:1} shows three different demand traces with single, two consecutive and three consecutive small peaks. For each trace, we run 100 simulations on different days and hours the probability of having the super-linear gain is given in Table 5.
\begin{table}[H]\label{Sec5:T4}
	\centering
	\begin{tabular}{|c|c|c|c|}\hline
		 & Single & Two consecutive  & Three consecutive\\
		\hline
		Probability & 0.56 & 0.64 & 0.76 \\
		\hline
	\end{tabular}
	\caption{Probability of super-linear gain happening in contiguous small triangular peaks}
\end{table}

Table 6 shows a clear positive correlation between the number of contiguous small peaks and the probability of winning the super-linear saving. When there is only one small peak, the probability of having the extra saving is 56\%. The probability will increase to 64\% if we have two such small peaks in a sequence, and to 76\% if we have three consecutive small peaks. This is very likely because that multiple consecutive small peaks creates a better opportunity for charging and discharging batteries aggreggately when optimizing for both regulation market and peak shaving.

\subsection{Key insights}
We observe the following three key characteristics of peak demand in achieving super-linear gain:
\begin{itemize}
\item flat peaks are more favored than sharp peaks for the joint-optimization method.
The randomness of regulation signal helps break down the one flat peak into several sharp peaks, thus we save more from doing peak shaving \emph{on the top of} providing regulation service. The super-linear saving comes from the joint-optimization framework, participating in one applications actually facilitates another application.
\item for a single small peak, which is low and narrow, could be regraded as a \textquotedblleft immediate state\textquotedblright\  between sharp and flat and has around 50\% to 60\% probability to see the non-linear saving.
\item if we have several such small peaks in a sequence, the probability of gaining the super-linear gain will gradually increase as it creates better battery chargning/discharging opportunity aggregately when optimizing two applications jointly.
\end{itemize}
In the next section, we will try to link our observations and analysis to the real data center power usage traces, and to evaluate the realistic value of the proposed joint-optimization framework.

\section{Experimental results}
\label{sec:eval}
In this section, we want to link our prior observations to the real data center power usage traces, and evaluate how the proposed joint-optimization framework works in practice. Here we present the result of a case study, where Appendix \ref{app:stat} presents the statistics about the peaks and valleys in data center power traces.

\subsection{Case study}
In order to evaluate how the proposed joint-optimization works in real world, we have obtained data from PJM Regulation D market and Microsoft geo-distributed data centers. Since the time resolution for the Microsoft data is 20s and the time resolution for regulation signal is 2s. We do sampling on PJM regulation data so that they have the same time resolution ($t_s=20s$). We use our optimization framework for each hour under four different policies: not using battery, using battery only for regulation service, using battery for peak shaving and the joint-optimization method. The electricity bills under four scenarios are denoted as $J$, $J^{r}$, $J^{p}$ and $J^{*}$ respectively. We use $J$, the original electricity bill when not using battery, as the reference to calculate the bill saving for other cases.

The total hours simulated is 4415 hours (half year). And the simulation results shows that the proposed joint-optimization framework could helps reduce the total electricity bill by 30\% compared with the original bill, which is quite a considerable amount of money saving each year for a MW size data center.

\begin{figure}
	\centering
	\includegraphics[width= 0.95 \columnwidth, height = 0.6 \columnwidth]{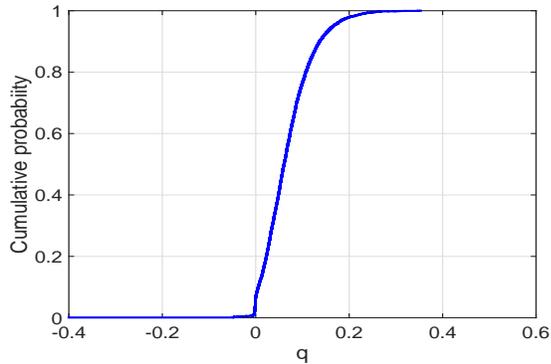}
	\caption{Cumulative probability distribution of super-linear benefits}
	\label{Sec5:P2:2}
\end{figure}

Further, we define a criteria $q$, which describes the super-linear saving ratio gained by the joint-optimization as (\ref{Sec6:P3:E1}),
\begin{equation}\label{Sec6:P3:E1}
q = \frac{(J-J^{*})-[(J-J^{r})+(J-J^{p})]}{J}\,,
\end{equation}
When $q$ is positive, it indicates that the saving from joint-optimization exceeds the sum of individually optimizing each component. And the larger $q$ is, the more extra saving we gain. If $q$ is negative, it suggests that we do not have the super-linear gain. Table 6 summarize the probability of winning the super-linear gain and the average ratio $q$. Fig.\ref{Sec5:P2:2} shows the cumulative probability distribution of $q$.

\begin{table}[H]
	\centering
	\begin{tabular}{|l|c|}\hline
		Total hours simulated & 4415 \\
		\hline
		Hours having super-linear gain & 4122 \\
		\hline
		Probability of having super-linear gain & 0.934 \\
		\hline
		Average super-linear saving ratio ($q$) & 0.065\\
		\hline
	\end{tabular}
	\caption{Summarization of the experiments}
	\label{Sec5:T1}
\end{table}

As shown in Fig.\ref{Sec5:P2:2} and Table \ref{Sec5:T1}, the joint optimization method, which optimizes the battery usage for regulation service and peak shaving at the same time, can win us extra saving compared with the \emph{sum} of each individual component. This experiment result matches our analysis and inference. Since the peaks of real data center traces are featured with triangular, low, narrow and continuity, they has a large probability of obtaining the super-linear gain. what's more, compared to prior works that consider single application for energy storage, our results also suggest that storage in data centers can have much larger impacts than previously thought possible.

\section{Conclusion}
\label{sec:con}
This paper addresses using battery storage in data centers to reduce their electricity bills. We consider two sources of cost savings: reducing the peak demand charge and gaining revenue from participating in regulation markets. We formulate a framework that jointly optimizes battery usage for both of these applications. Surprisingly, we observe that a superlinear gain can often be obtained: the saving from the joint optimization can be larger than the sum of the individual savings from devoting the battery to one of the applications. Using synthetic and real data center load traces, we perform statistical analysis of the nature of the gain. Our simulation show that a data center in the PJM control area can reduce its cost by 30\% if its battery is optimally used.

Optimization problems in this paper is solved in an offline manner, which is not practical since the regulation signal is generated by system operators and is unknown in advance. Developing online control strategies is a major direction of our future work. 

\section{Acknowledgements}
The work of Y. Shi, B. Xu and B. Zhang are partially supported by the University of Washington Clean Energy Institute.

\bibliographystyle{abbrv}
\bibliography{sigproc,dc_bib}  

\appendix
\label{app:stat}
\subsection{Statistics of real life data-center traces}
In the prior section, we propose a peak abstraction method in terms of peak height, peak width and peak shape. We use the abstraction method to characterize peaks into eight categories and studied the effectiveness of the joint-optimization method for each category. We apply the same methodology to the real data center power demand and characterize its peaks in terms of peak height, width and shape. We then perform statistical analysis and classify them into one of the eight categories.

\subsubsection{Setup}
We use power measurement data from Microsoft data center over a half year period. The time resolution for the raw data is 20 seconds. Here we set $f$, the percentage of power capping to 20\%, and calculate the peak threshold as $C_f = (1-f) \times d + p_{min}$. Notice that $C_f$, $d$ and $p_{min}$ are calculated by each day. Once setting the threshold $C_f$, the original demand series could be divided into several peak intervals and valley internals. And we want to figure out the statistical distribution of the peak height, peak width, peak shape of all the peak intervals.

\subsubsection{Statistical results}
\begin{figure}[ht]
	\centering
	\begin{subfigure}[b]{0.475\columnwidth}
		\centering
		\includegraphics[width=\columnwidth]{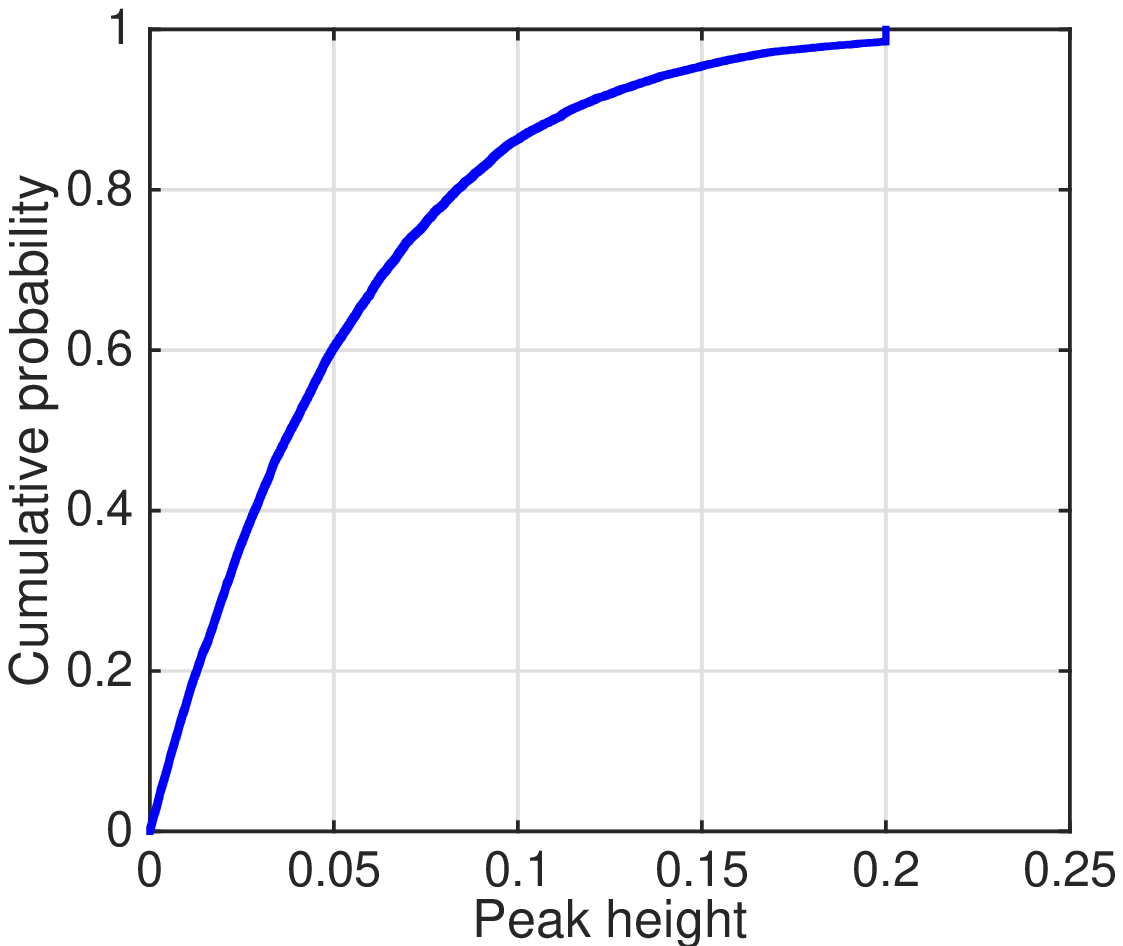}
		\caption[Network2]%
		{{\small Cumulative probability distribution of peak height}}
		\label{Sec5:P1:1}
	\end{subfigure}
	\hfill
	\begin{subfigure}[b]{0.475\columnwidth}
		\centering
		\includegraphics[width=\columnwidth]{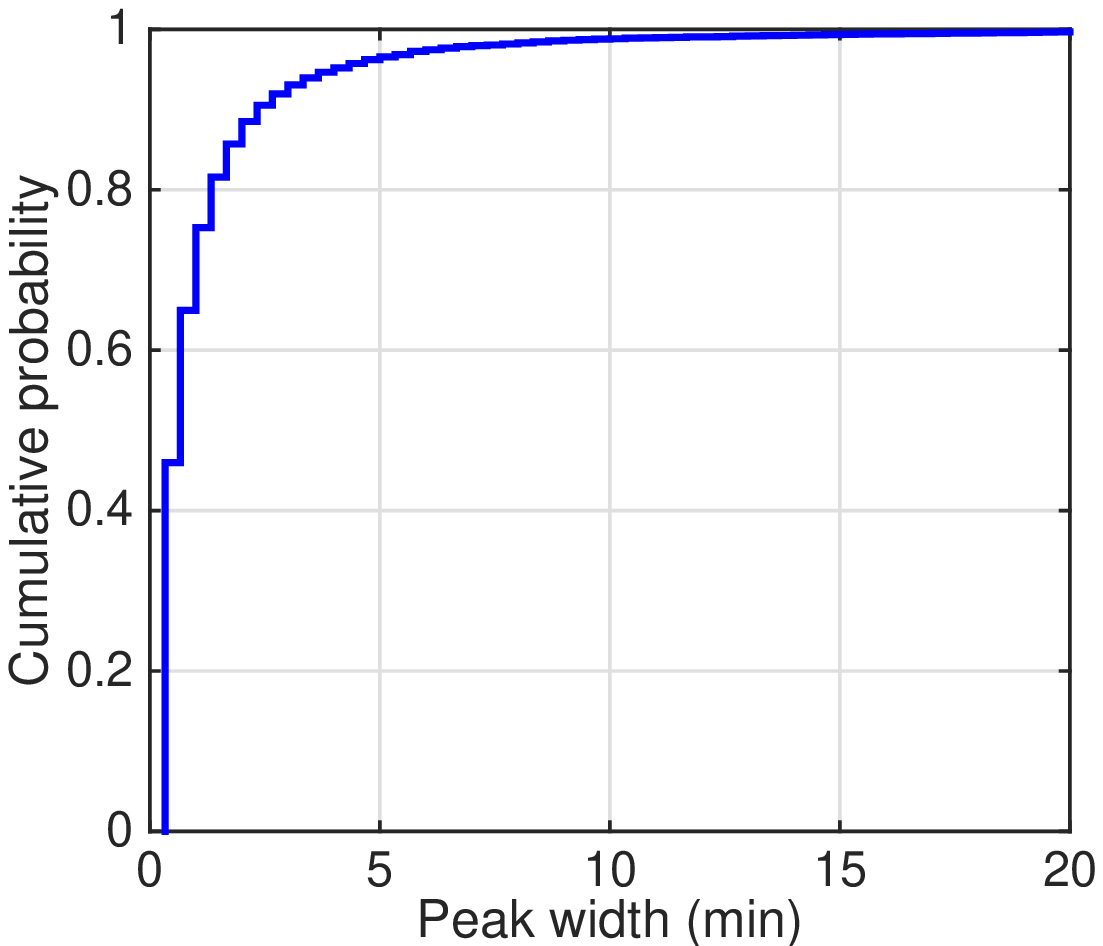}
		\caption[]%
		{{\small  Cumulative probability distribution of peak width}}
		\label{Sec5:P1:2}
	\end{subfigure}
	\vskip\baselineskip
	\begin{subfigure}[b]{0.475\columnwidth}
		\centering
		\includegraphics[width=\columnwidth]{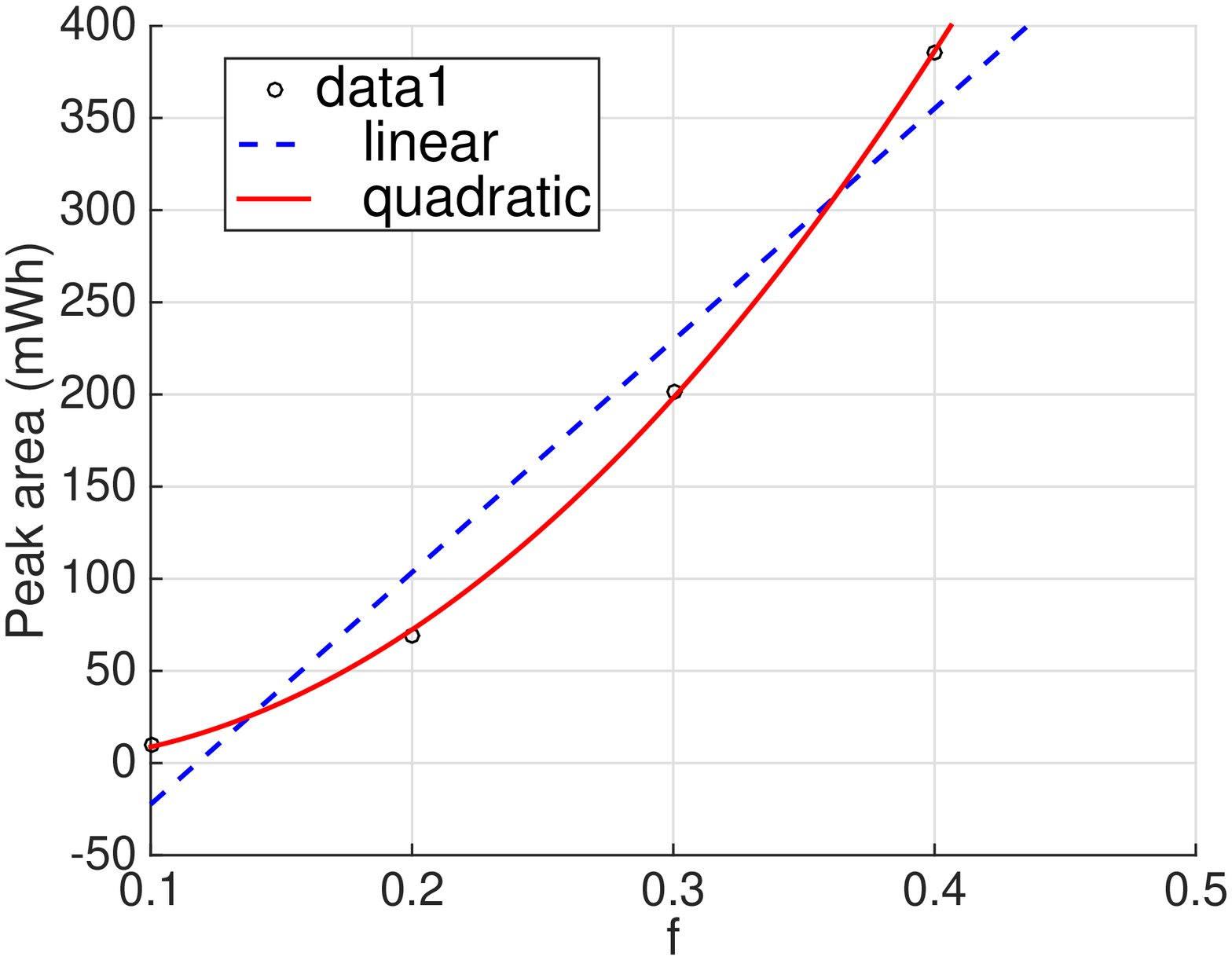}
		\caption[]%
		{{\small Shaved area VS different capping fraction $f$}}
		\label{Sec5:P1:3}
	\end{subfigure}
	\quad
	\begin{subfigure}[b]{0.475\columnwidth}
		\centering
		\includegraphics[width=\columnwidth]{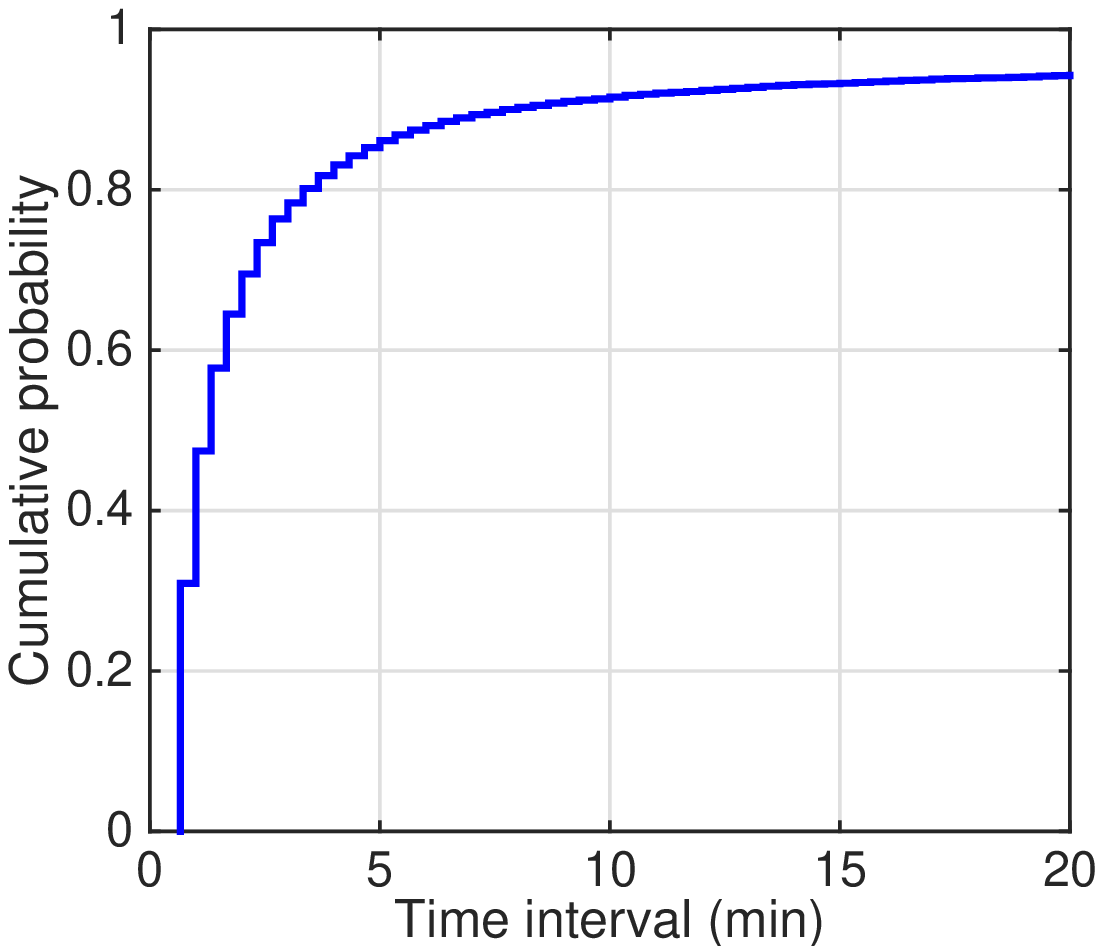}
		\caption[]%
		{{\small  Cumulative probability distribution of time interval between peaks}}
		\label{Sec5:P1:4}
	\end{subfigure}
	\caption[ The average and standard deviation of critical parameters ]
	{\small Statistical analysis for real life data center}
	\label{Sec5:P1:s1}
\end{figure}

(1) \textbf{Peak height:} Fig. \ref{Sec5:P1:1} depicts the probability distribution of peak height. Remember that we set the power capping fraction $f$ to 0.2. So all the peak height is measured between $0$ to $0.2$ depending on the p$_{max}$ over that peak interval:
\begin{equation*}
PH = \frac{max_{t_a \leq t \leq t_b}{[s(t)-C_f]}}{d}\,,
\end{equation*}
For example, if the $PH_i$ equal to $0.1$, it means the height of peak $i$ is half tall as the highest peak of the day. If $PH_i$ equal to $0.05$, it represents a even lower peak, only a quarter as high. While if the peak height is close to $0.2$, it indicates a tall peak close to the highest peak of the day. The results given in Fig. \ref{Sec5:P1:1} show that nearly 90\% of the peaks are half-height as the highest peak of the day. What's more, 60\% of the peaks are only a quarter height as the highest peak. This results suggest most of the peaks are \emph{low} in terms of height.

(2) \textbf{Peak width:} Fig. \ref{Sec5:P1:2} gives the cumulative probability distribution of the peak width. We could find most real life data center peaks are quite \emph{narrow}, among which 90\% of the peaks last less than 2 min and 75\% less than 1 min.

(3) \textbf{Peak shape:} In the prior section, we defined the peak shape, as one of the attributes to abstract the peaks. In terms of shapes, we roughly classify peaks into two categories: rectangular peaks and triangular peaks. Different peak shapes may influence the super-linear saving even for the same peak width and peak height. In order to find out whether the real data center peaks are more similar to triangular or to rectangular, we consider a method based on the change rate of peak area.

\begin{figure}[ht]
	\centering
	\includegraphics[width= \columnwidth, height= 0.618 \columnwidth]{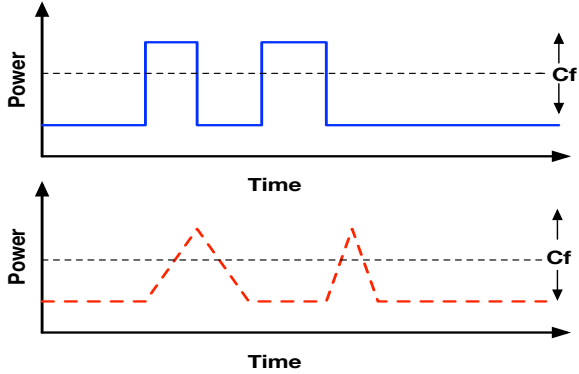}
	\caption{Change of peak area for rectangular shape and triangular shape peak}
	\label{Sec5:P1:5}
\end{figure}

As shown in Fig. \ref{Sec5:P1:5}, we find that the peak area (total area above the threshold $C_f$) changes differently for rectangular and triangular peaks. For a rectangular peak, the peak area increases in a linear manner while $C_f$ moves down; while for triangular peaks, the peak area increases in a quadratically manner when $C_f$ moves down. Given this property, we calculate the peak area with respect to different capping fraction $f$ (corresponding to different $C_f$), and use linear as well as as quadratic model to fit. Fig.\ref{Sec5:P1:3} shows the quadratically model fits perfectly for $f$-peak area pair. Therefore, we could infer that most of the real data center peaks are \emph{triangular}.

(4) \textbf{Consecutive peaks:} In order to provide a more detailed understanding of the correlation between peaks, we have performed statistical analysis on the time intervals between peaks. Fig. \ref{Sec5:P1:4} shows that there is a good continuity between peaks. 60\% of the peaks happens within 2 minutes after the prior one.

To summarize the above experiment and statistics analysis results, there are three characterizations about the peaks of Microsoft data center power demand traces:
\begin{itemize}
	\item Most of the demand peaks are low with short duration;
	\item Most of the peaks tend to have a triangular shape;
	\item There is a high probability that peaks happen contiguously.
\end{itemize}
There features characterizes the Microsoft trace as the \emph{contiguous small peaks} type. Based on the analysis in Section 4, we conclude that there should bes a large probability of winning the super-linear gain when implementing the joint-optimization method in real world data center like Microsoft's.

\end{document}